\documentclass[
    ,final            % use final for the camera ready runs
 ,  aas_macros                 % add further options here if necessary
  ]
  {aipproc}

\layoutstyle{6x9}

\usepackage{natbib}

\begin{document}

\title{Post-Main Sequence Evolution of Debris Discs}

\author{Amy Bonsor}{
  address={Institute of Astronomy, University of Cambridge, Madingley Road,
  Cambridge CB3 0HA, UK}
}

\author{Mark C. Wyatt}{
  address={Institute of Astronomy, University of Cambridge, Madingley Road,
  Cambridge CB3 0HA, UK}
}


\begin{abstract}
 The population of debris discs on the main sequence is well constrained, however very little is known about debris discs around evolved stars. In this work we provide a theoretical framework that considers the effects of stellar evolution on debris discs; firstly considering the evolution of an individual disc from the main sequence through to the white dwarf phase, then extending this to the known population of debris discs around main sequence A stars. It is found that discs around evolved stars are harder to detect than on the main sequence. In the context of our models discs should be detectable with Herschel or Alma on the giant branch, subject to the uncertain effect of sublimation on the discs. The best chances are for hot young white dwarfs, fitting nicely with the observations e.g the helix nebula \citep{helix} and 9 systems presented by Chu \& Bilikova (this volume). Although our baseline models do not predict such a high rate of detectable discs.

\end{abstract}

\maketitle

%%%%%%%%%%%%%%%%%%%%%%%%%%%%%%%%%%%%%%%%%%%%
%% MAINMATTER
%%%%%%%%%%%%%%%%%%%%%%%%%%%%%%%%%%%%%%%%%%%%

\section{Introduction}

The first dusty disc around a main sequence star was observed in 1984 around Vega \citep{vega1984}. Since then our knowledge of such systems has improved significantly, and it is now known that 32\% of A stars exhibit excess emission in the infrared, over and above the stellar photosphere \citep{su06}. This is thermal emission from dust particles orbiting the star in a debris disc. Debris discs are collisionally dominated in that the smallest bodies in the system are continuously replenished by collisions between larger objects and are subsequently removed by radiation pressure. The long term evolution of such systems can be modelled by the steady state collisional models of \cite{wyatt07} and is expected to be a slow decline in brightness as the disk mass is depleted due to collisional erosion. A decrease in brightness with age is indeed observed \citep[e.g.][]{su06} and can be well fitted with the models of \cite{wyatt07}, allowing such models to characterize the population of main sequence A stars debris discs reasonably accurately.

Dust is also seen around some post-main sequence stars. In some cases this dust can be a result of the evolution of the star, for example material emitted in the stellar wind form spherical shells of dust that are observed around AGB stars (e.g. \citealt{AGBshell2010}) or even stable discs observed around post-AGB stars, possibly linked to binarity (e.g.~\citealt{winkel09postAGBbinary}). Infrared excess observed around giant stars, e.g. \cite{jura99}, and the helix nebula \citep{helix}, on the other hand, has been interpreted as a disc similar to debris discs on the main sequence (although alternative interpretations exist, see e.g. \citealt{kimzuckerman01}). Hot dust is also observed in small radii ($<$0.01AU) discs around white dwarfs, e.g. \cite{farihi09} or \cite{2007Kuchner}, again inferred to originate from a debris disc. However, in contrast to main sequence debris discs, these discs cannot be in steady state since material at such small radii has a short lifetime. Rather models suggest that these discs are formed when an asteroid approaches close to the star where it is tidally disrupted \citep{jurasmallasteroid}.

There are not yet enough observations of discs around post-main sequence stars to fully understand the population and it is not clear how the few discs that have been discovered around post-main sequence stars relate to the progenitor population of debris discs on the main sequence. In this contribution we discuss the evolution of debris discs beyond the main sequence, with particular reference to models presented in \cite{bonsor10}. These models intend to be a theoretical framework in which all of the effects of stellar evolution on a disc are investigated. We use these models to evolve the population of debris discs observed around main sequence A stars \citep{rieke05} \citep{su06} and discuss the detectability of such discs in terms of the observations of dust around evolved stars.

\begin{figure}
\includegraphics[width=0.5\textwidth] {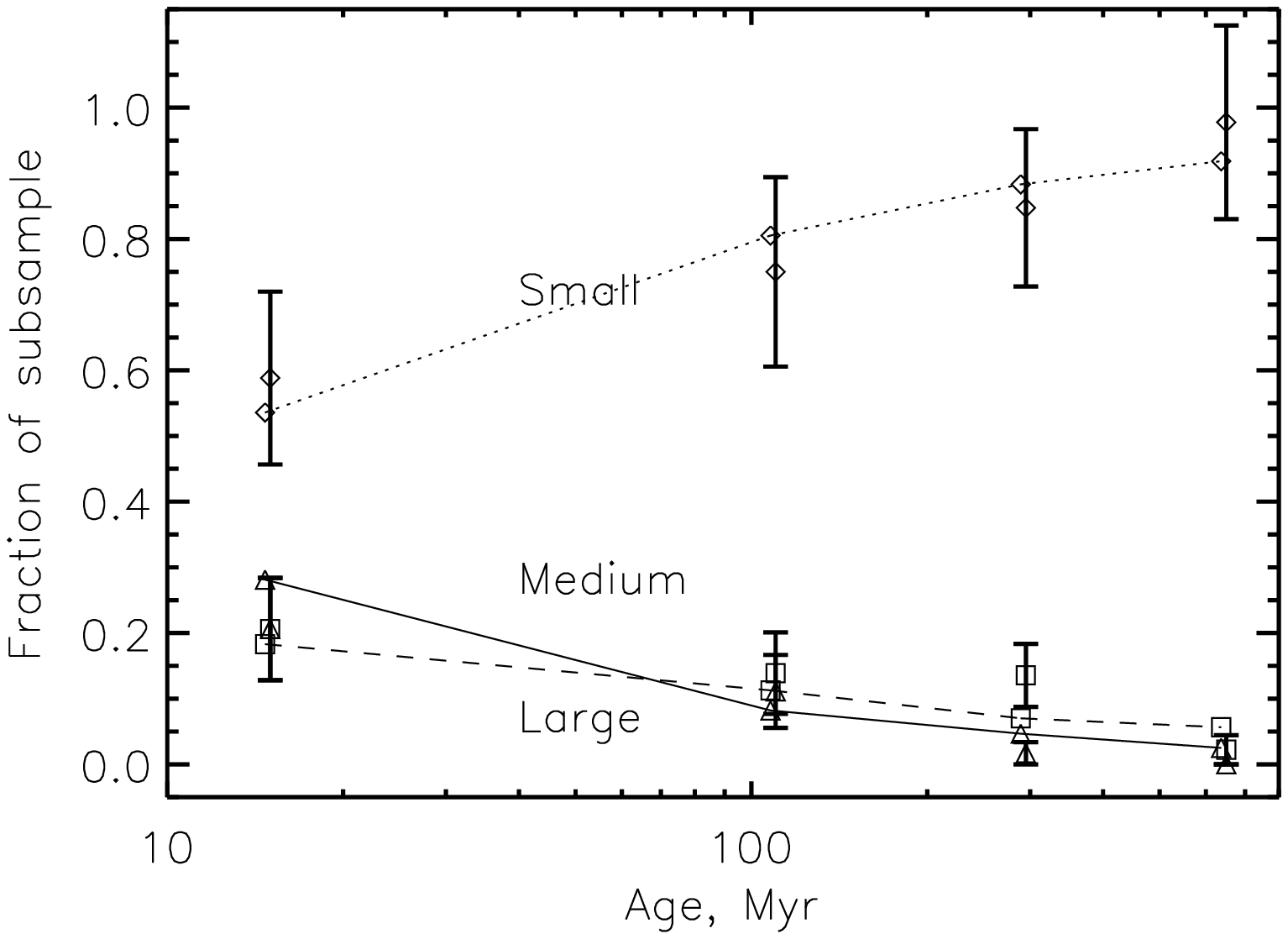}
\includegraphics[width=0.5\textwidth] {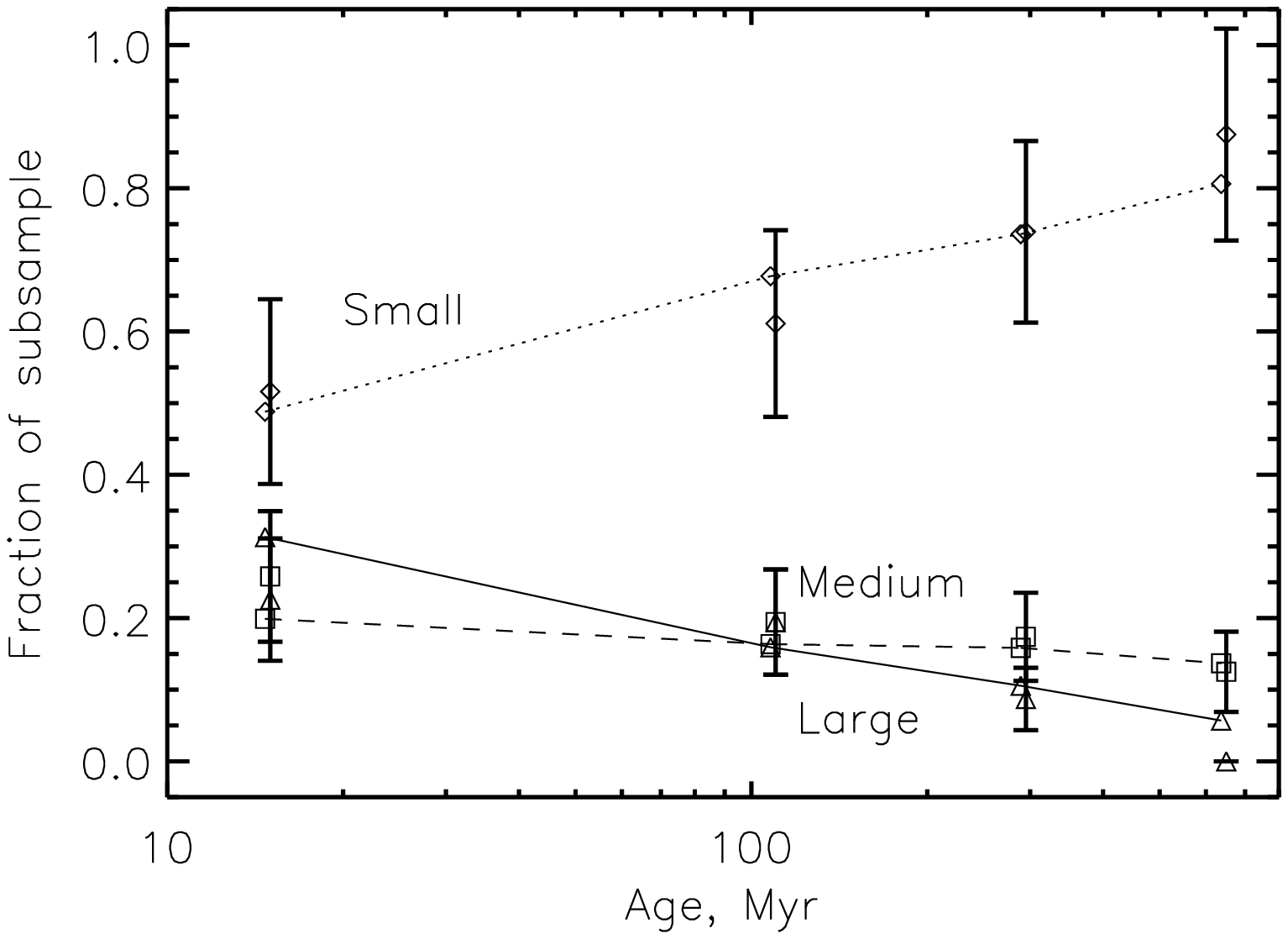}
\caption{The fit to the observations of \citet{su06} at 24 (left) and 70$\mu$m (right), comparable to Fig.2 of \citet{wyatt07}. The plots show the fraction of stars with flux ratios in different age bins ($<$30Myr, 30-190Myr, 190-400Myr), at 24$\mu$m $\frac{F_{disc}}{F_*}=$ 1-1.25 (diamond: small excess), 1.25-2 (square: medium excess), $>$2 (triangle: large excess) and similarly at 70$\mu$m $\frac{F_{disc}}{F_*}$= 1-5 (diamond:small excess), 5-20 (square: medium excess), $>$20 (triangle: large excess). Observed values are shown with $\sqrt{N}$ error bars, whilst model values are joined with dotted, dash and solid lines, for small, medium and large excess.}
\label{fig:fit}
\end{figure}

\section{Population of debris discs around main sequence stars}
The population of debris discs around main sequence stars is relatively well constrained. For our models we consider Spitzer observations at 24 and 70$\mu$m  around A stars. These were modelled in detail using the steady state collisional models of \cite{wyatt07}. The observations were split into stars of different ages and small, medium and large excess. A population model reproduces these observations very well, see Fig.~\ref{fig:fit} In this work the models of \cite{wyatt07} were extended to include emission properties of realistic grains.

\section{Models for post-main sequence evolution of discs}
This population of discs from the main sequence is combined with models for the evolution of the star from \cite{sse} to determine what the population of debris discs around evolved stars should look like. Further details of these models can be found in \cite{bonsor10}.

\begin{figure}
\includegraphics[width=0.5\textwidth]{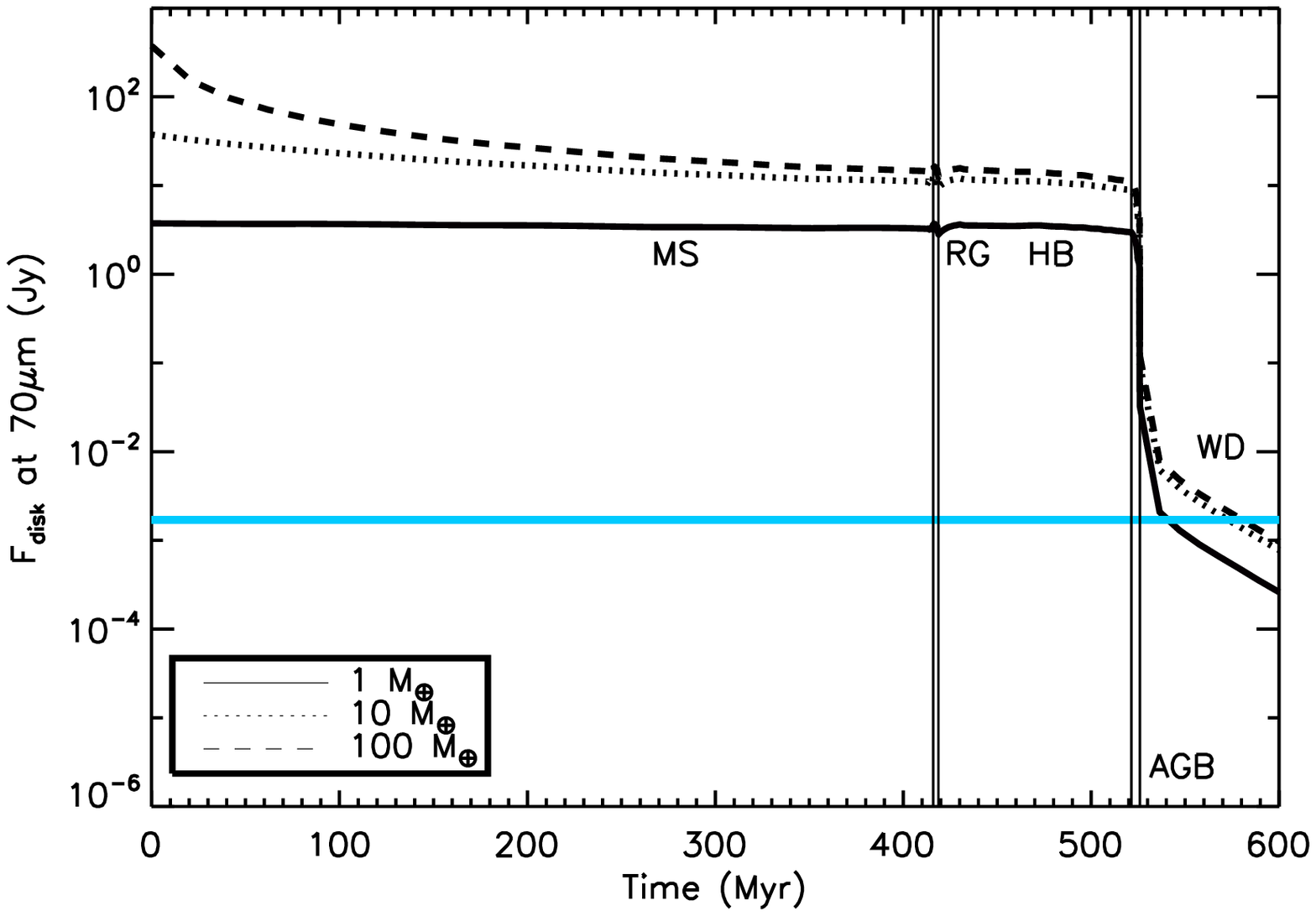}
\includegraphics[width=0.5\textwidth]{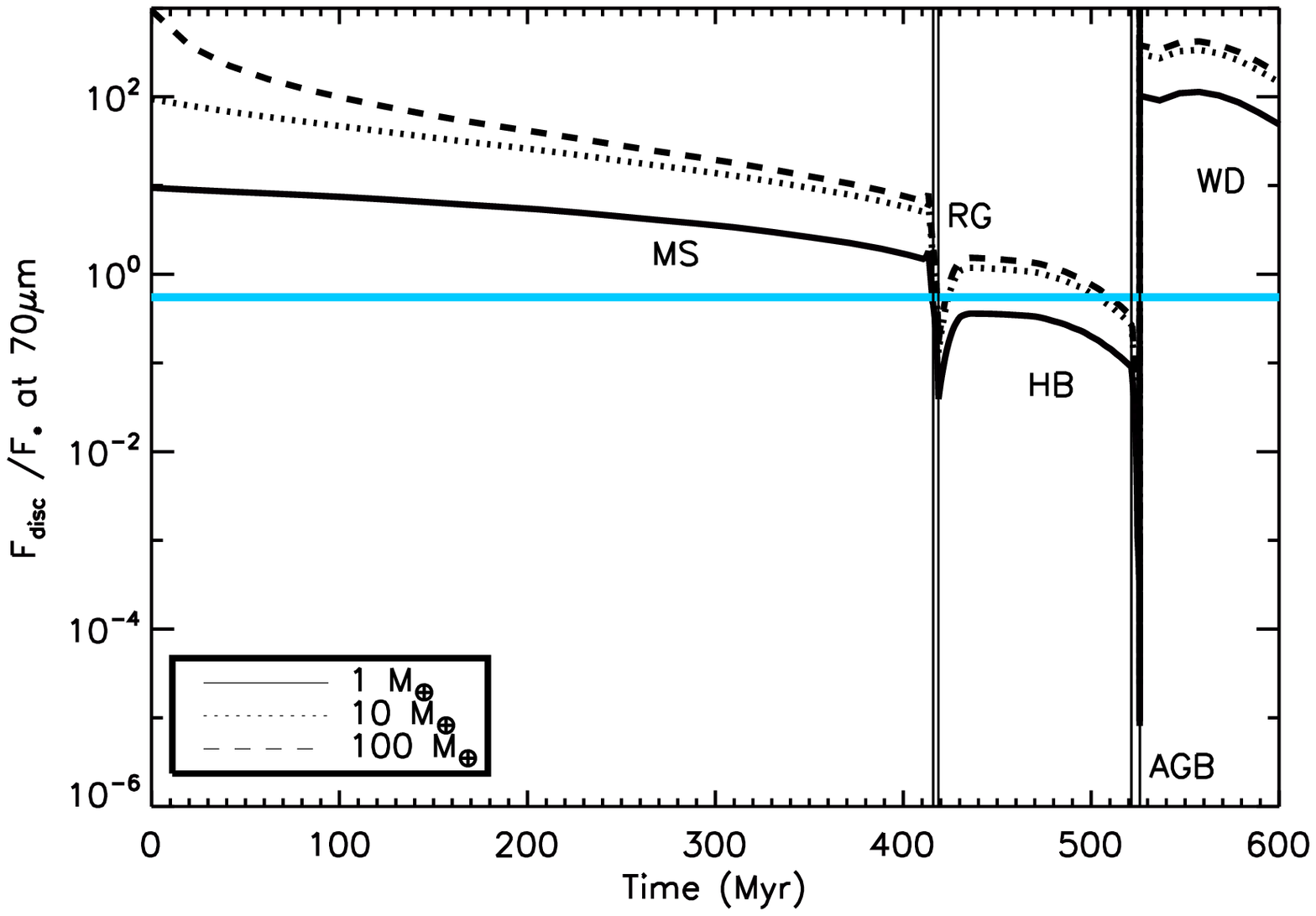}

\caption{The evolution of the total flux from the disc (left panel) and the ratio of the flux from the disc to the flux from the star (right panel) at $70\mu$m, as the star evolves. The thick gray line in the left plot is the sensitivity limit of 110$\mu$Jy, whilst in the right plot it shows the calibration limit of $R_{\mathrm{lim}}=0.1$, for Spitzer at $70\mu$m. The star is a 2.9$M_{\odot}$ star, with solar metallicity ($Z=0.02$), at 10pc and the disc has an initial radius of 100AU.}
\label{fig:f_R}
\end{figure}

The two changes to the star that are most important for the disc are the changes in stellar luminosity and mass. Stellar luminosity remains fairly constant on the main sequence. However it increases by several orders of magnitude on the giant and asymptotic giant branch and drops by several orders of magnitude for white dwarfs. The disc temperature follows this luminosity evolution. The stellar mass, on the other hand remains fairly constant until the end of the asymptotic giant branch where it drops by a factor of 2 or 3. This mass loss is adiabatic and critically important for white dwarf discs as it means that they have large radii.

Collisions dominated the discs evolution on the main sequence and this continues post-main sequence. The mass in the disc is reduced as collisions grind the largest bodies down into smaller and smaller dust particles. The smallest particles are removed from the system. On the main sequence it is radiation pressure that blows small grains out of a debris disc. However post-main sequence there is a competition between 5 processes. Not only do radiative forces blow small grains out of the disc, but they causes them to spiral into the star, under PR-drag. This is important for small radii discs on the main sequence. The stellar wind is also important post-main sequence, especially during the high mass loss phase on the asymptotic giant branch. Similarly to radiative forces, the stellar wind can either cause grains to become unbound or spiral into the star under stellar wind drag. Radiation pressure removes the largest grains, apart from during the high mass loss phase at the end of the AGB when stellar wind drag dominates. 
Small grains can also sublimate, silicates survive the entire evolution, whilst water ices particles are removed. The effect of sublimation is not fully understood at present, and will be discussed further later.

All of these effects can be put together to determine the evolution of an individual disc. Fig.~\ref{fig:f_R} shows this for an example disc with a radius of 100AU, initial mass of 30M$_{\oplus}$, around a 2.9M$_{\odot}$ star at a distance of 10pc,  observed with Spitzer at 70$\mu$m. The flux from the disc decreases along the main sequence as the mass in the disc is depleted by collisions. The disc flux doesn't increase on the giant branch, despite the increase in disc temperature. This is because small grains which have the largest cross-sectional area for emission are removed by radiation pressure. The disc mass and hence flux continue to decrease on the horizontal branch, remain relatively constant on the AGB, since here even larger grains are removed by radiation pressure. The disc flux falls dramatically with the stellar luminosity as the star becomes a white dwarf.

In order to detect a disc we require that not only the emission from the disc bright enough to be detected from earth, i.e. the disc flux is above the sensitivity of the instrument, but that the disc can be detected over and above the stellar photosphere. This requires that the ratio of the disc flux to the stellar flux is greater than the calibration limit. Fig.~\ref{fig:f_R} shows the ratio of the disc flux to the stellar flux for the same example disc, compared to the calibration limit of Spitzer at 70$\mu$m (0.55). This example disc can be detected with Spitzer at 70$\mu$m both early on the giant branch and early in the white dwarf phase. It was found to be generically true for all discs in our population, that a proportion are detectable on the main sequence, this decreases on the giant branch, still further on the horizontal branch and only discs around very nearby hot white dwarfs are detectable.

\begin{table}
\centering
\begin{minipage}{0.6\textwidth}
\centering                          % centering table
\begin{tabular}{ccr}          % creating eight columns
\hline
&  &  \tablehead{1}{r}{b}{Giant Branch}    \\ [-1.0ex]
\tablehead{1}{c}{b}{Instruments}& 
\tablehead{1}{c}{b}{Sensitivity (mJy)} &    
\tablehead{1}{r}{b}{d$\mathbf{< 100}$pc} \\

\hline                               % inserts single-line
& &  \tablehead{1}{r}{b}{$\mathbf{\%}$}   \\

\hline 

IRAS at 60$\mu$m$^{\dag}$ &100$^{**}$& 1.7 \\
Spitzer at 24$\mu$m& 0.11 $^{\ddag}$  & 14.0  \\
Spitzer at 70$\mu$m&14.4$^{\ddag}$  &9.3 \\
Spitzer at 160$\mu$m & 40 $^{\ddag}$   &4.2 \\
Herschel PACS at 70$\mu$m & 4$^{\S}$  &9.6 \\
Herschel PACS at 160$\mu$m & 4$^{\S}$  &12.2 \\
Herschel SPIRE at 250$\mu$m & 1.8$^{\S}$ &12.8 \\
Herschel SPIRE at 350$\mu$m & 2.2$^{\S}$  &10.8 \\
Alma at 450$\mu$m  & 80$^{\P}$ &7.0 \\
Alma at 1.2mm &0.25& 2.2 \\
Spica at 200$\mu$m & 0.1$^{||}$  &12.0 \\

\hline
No. of stars$^{\dag \dag}$ &  &  1050$^{\dag}$ \\
 
\hline
%\footnotetext[b] {The calibration limit of Spitzer at 70$\mu$m. }
%\footnotetext[2]{ Hypothetical calculations for a reduction of an order of magnitude in the calibration limit.}  
\footnotetext[2]{Only stars with magnitudes brighter than 4.0 are considered such that the sample can be compared with ~\cite{jura90}}
\footnotetext[3]{http://irsa.ipac.caltech.edu/IRASdocs/iras\_mission.html}
\footnotetext[4]{ \cite{wyattreview}} 
\footnotetext[5]{http://herschel.esac.esa.int/science\_instruments.shtml } 
\footnotetext[6]{http://www.eso.org/sci/facilities/alma/observing/specifications/}
\footnotetext[7]{ \cite{spica}}
\footnotetext[8]{The number of evolved A stars, calculated from the space density of A stars \citep{phillips09}}
                         % inserts single-line
\end{tabular}
\end{minipage}
\caption{Detection of discs around evolved stars}
\label{tab:percent}
\end{table}

\section{Discs around giant stars}
Using our population models \citep{bonsor10} the percentage of the evolved population of main sequence A star discs that have detectable excess on the giant branch can be calculated and is shown in Table~\ref{tab:percent}. The figures for IRAS are inline with previous observations, whilst the figures for Spitzer and Herschel are significantly less than on the main sequence. There is however, one caveat, for these observations which is the uncertain effect of sublimation on the disc. If sublimation causes the release of a population of small silicate grains, as in the models of \cite{juraotherkb}, then the number of discs that are detectable will increase. On the other hand if it truncates the collisional cascade at larger grain sizes, the disc flux and hence number of detectable discs will decrease. This gives future observations of giant stars with Herschel the potential to constrain the effect of sublimation on debris discs, and hence is of great interest.

\section{Discs around white dwarfs}

\label{sec:wdobs}
As can be seen in Fig.~\ref{fig:f_R} the disc flux falls off rapidly as the star cools during the white dwarf phase. It is therefore very hard to detect debris discs around white dwarfs. Observations of debris discs around white dwarfs in our baseline model are sensitivity limited and only the most massive discs around the closest, youngest white dwarfs are detectable. 

\begin{figure}[t]
\includegraphics[width=0.85\textwidth]{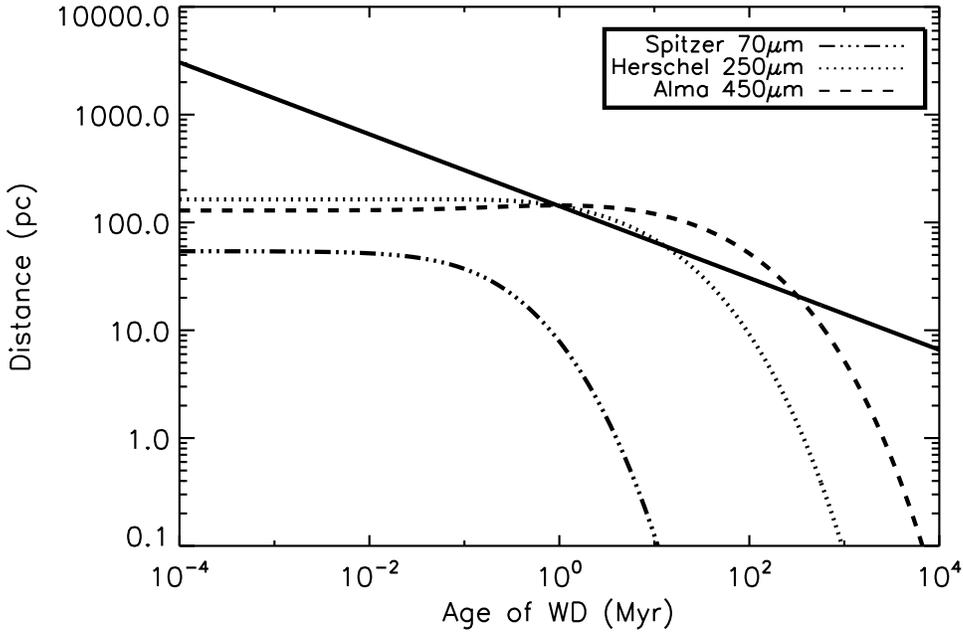}
\caption{The dash-dotted, dotted and dashed lines show the maximum distance out to which a disc initially at 100AU, with a mass of $10M_{\oplus}$, around an evolved 3.8$M_{\odot}$ white dwarf, of a given age can be detected with Spitzer at 70$\mu$m, Herschel at 250$\mu$m and Alma at 450$\mu$m, respectively, whilst the solid line shows the distance within which there is one white dwarf younger than the given age, calculated using the space density of A stars from \cite{phillips09}.}
\label{fig:wdmaxdist}
\end{figure}

There is, however, a balance between young white dwarfs being the most luminous and therefore having the brightest discs and the low volume density of young white dwarfs such that they are more likely to be found at greater distances from the sun. Fig.~\ref{fig:wdmaxdist} shows the maximum distance out to which discs around white dwarfs can be detected as a function of cooling age, for discs at 100AU with Spitzer at 70$\mu$m, Herschel SPIRE at 250$\mu$m and Alma at 450$\mu$m. This is compared to the distance within which one white dwarf of a given cooling age is found, according to the space densities of \cite{phillips09}. The maximum distance out to which discs can be detected is never significantly greater than the distance within which there is one white dwarf and it is therefore unlikely that such a system can be observed. There is an optimum cooling age for detecting white dwarf discs, which varies with wavelength, for Spitzer at 70$\mu$m it is $\sim$1Myr, whilst for Herschel SPIRE at 250$\mu$m it is $\sim$10Myr and Alma at 450$\mu$m $\sim$100Myr. As the disc temperature drops, the disc flux decreases, more rapidly at the shorter wavelengths. This means that for a young population of white dwarfs, the best chances of detecting debris discs are at the shorter wavelengths of Spitzer or Herschel, whilst for a sample that includes older stars Alma would be better. However, overall, the best chances of detecting such a system are with the longer wavelengths of Herschel or Alma. 

Focusing on Spitzer at 70$\mu$m, if for some reason our models under-predicted the flux from (or mass in) such discs by approximately an order of magnitude a disc would be most likely to be detected around a white dwarf of less than 5Myr old at a distance of around 200pc. The only detection of excess around a white dwarf that resembles a main sequence debris disc is the helix nebula \citep{helix}, a young white dwarf with a cooling age significantly less than 5Myr, surrounded by a planetary nebula at 219pc. This fits nicely with our models, especially given that alternative explanations that increase the disc flux exist, for example the trapping of bodies in resonances \citep{dong10}.

\begin{figure}[t]
\includegraphics[width=0.82\textwidth]{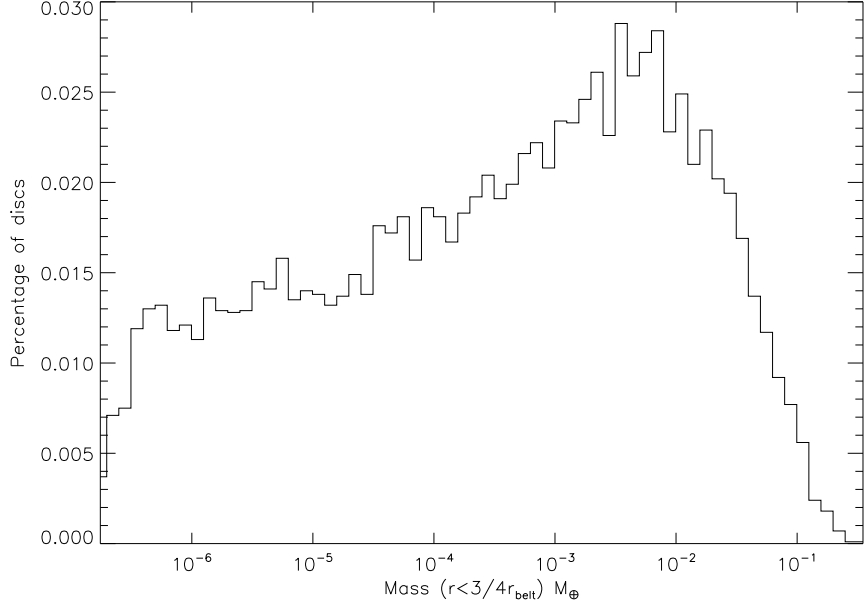}
\caption{A histogram showing the amount of mass left inside of the main belt (between r=0 and $r=\frac{3}{4} r_{belt}$) at the end of the AGB for the population of discs in our models.}
\label{fig:hist}
\end{figure}

These low probabilities for detecting debris discs around white dwarfs fits with the fact that Spitzer observations of white dwarfs that have only found one white dwarf with infrared excess fitted by a disc with a radius of the same order of magnitude of main sequence debris discs. There are however $\sim$20 observations of hot, dusty discs around white dwarfs that are best fitted by discs of radii on the order the solar radius e.g. \cite{farihi09}, \cite{reach05}.  \cite{farihi09} estimate that 1-3$\%$ of white dwarfs with cooling ages less than 0.5 Gyr possess hot IR excess. The minimum radius of a disc in our population is $\sim$10AU and therefore these observations cannot be explained by the discs in our population. Material in discs with such a small radius will have a very short lifetime and must, therefore, be replenished. Within the context of the current model we have identified a potential source of material for such discs. Stellar wind drag was included in the current models in as far as it truncates the collisional cascade on the AGB. Material that leaves the disc will spiral in towards the sun, most of it being accreted onto the star during the AGB, however some mass will be left between the inner edge of the belt and the star, at the end of the AGB. Fig.~\ref{fig:hist} shows this mass for all the discs in our population. The masses in Fig.~\ref{fig:hist} are significantly higher than the typical dust masses for these hot discs e.g. $3.3 \times 10^{-10} M_\oplus$ of GD166-58 \citep{farihiI}, and there are even a significant proportion of the population with more mass than the largest such disc, GD362, with a mass of 0.017$M_{\oplus}$ \citep{juraxray}. However a mechanism is still required to move this material in closer to the star. This could potentially be scattering by planets inside of the disc or the dynamical effects of mass loss on the disc. The effects will be considered in more detail in a future piece of work.

\section{Conclusions}
\label{sec:conclusions}

We have presented a piece of that work \citep{bonsor10} that provides a theoretical framework in which all of the effects of stellar evolution on debris discs are considered. Our models find that it is significantly harder to detect debris discs around evolved stars. The fraction of discs with detectable excess decreases significantly on the giant branch, yet further on the horizontal branch and discs around white dwarfs are very hard to detect.  

Approximately 10\% of giant stars should have detectable discs, whilst the highest chances of detecting a debris disc around a white dwarf are for very young, nearby white dwarfs. This fits with the absence of such detections apart from the helix nebula \cite{helix} and a handful of other systems also presented in this volume (see entries by Chu and Bilikova). According to our models the helix nebula, although not anticipated to have a detectable disc is at the optimum age and distance for such a detection, should for some reason our models underpredict the disc flux. 

Evolved main sequence debris discs cannot directly explain the near-IR observations of hot dusty discs around white dwarfs e.g. \cite{farihi09}. The minimum disc radius in our population is 10AU, whilst these sytems have radii less than R$_{\odot}$. However the fact that we predict a population of cold, undetectable planetesimal belts around white dwarfs has great implications for such discs. These belts could potentially provide the reservoir of material that replenishes such systems.

%%%%%%%%%%%%%%%%%%%%%%%%%%%%%%%%%%%%%%%%%%%%%%%%
%% BACKMATTER
%%%%%%%%%%%%%%%%%%%%%%%%%%%%%%%%%%%%%%%%%%%%%%%%

\begin{theacknowledgments}
I would like to acknowledge the support of an STFC funded PhD (Bonsor) and various fruitful discussions with colleagues. 

\end{theacknowledgments}


 \bibliographystyle{aipproc}   % if natbib is available
%%% %\bibliographystyle{aipprocl} % if natbib is missing

 \bibliography{ref}

\IfFileExists{\jobname.bbl}{}
 {\typeout{}
  \typeout{******************************************}
  \typeout{** Please run "bibtex \jobname" to optain}
  \typeout{** the bibliography and then re-run LaTeX}
  \typeout{** twice to fix the references!}
  \typeout{******************************************}
  \typeout{}
 }

\end{document}